# A microscopic cranking model for nuclear collective rotation
# I: rigid-plus-irrotational-flow rotating frame

P. Gulshani

NUTECH Services, 3313 Fenwick Crescent, Mississauga, Ontario, Canada L5L 5N1
Tel. #: 647-975-8233; matlap@bell.net

We derive in a simple manner and from first principles the Inglis's semi-classical phenomenological cranking model for nuclear collective rotation. The derivation transforms the nuclear Schrodinger equation (instead of the Hamiltonian) to a rotating frame using a product wavefunction and imposing no constraints on either the wavefunction or the nucleon motion. The difference from Inglis's model is that the frame rotation is driven by the motions of the nucleons and not externally. Consequently, the transformed Schrodinger equation is time-reversal invariant, and the total angular momentum is the sum of those of the intrinsic system and rotating frame. In this article, we choose the rotation of the frame to be given by a combination of rigid and irrotational flows. The dynamic angular velocity of the rotating frame is determined by the angular momentum of the frame and by a moment of inertia that is determined by the nature of the flow combination. The intrinsic-system and rotating-frame angular momenta emerge to have opposite signs. The angular momentum of the rotating frame is determined from requiring the expectation of the total angular momentum to have a given value. The transformed Schrodinger equation has, in addition to the Coriolis energy term, a rigid-flow type kinetic energy term that is absent from the conventional cranking model Schrodinger equation. Ignoring the relatively small effect of the fluctuations in the angular velocity and for a self-consistent deformed harmonic oscillator mean-field potential, the resulting Schrodinger equation is solved for the ground-state rotational band excitation energy and quadrupole moment in different configurations of $^{20}_{10}Ne$ and the results are compared with those of the conventional cranking model and empirical data.

## 1. Introduction

The self-consistent cranking model [1,2] is frequently used to study collective rotational properties of deformed nuclei [3-26 and references therein]. The model assumes that the anisotropic nuclear potential $V$ is rotating at a constant angular frequency $\omega_{cr}$ about $x$ or 1 axis. The model time-dependent Schrodinger equation:

$$i\hbar \frac{\partial}{\partial t}|\Psi_{cr}\rangle = H_{cr}|\Psi_{cr}\rangle \qquad (1)$$

where:

$$H_{cr} \equiv \frac{1}{2M} \sum_{n,j=1}^{N,3} p_{nj}^2 + V_{cr}(\vec{r}_n), \quad \vec{r}_n = R(\omega_{cr} t) \vec{r}_n' \qquad (2)$$

and $R$ is an orthogonal matrix and $\vec{r}_n'$ is the particle coordinates relative to the rotating frame, is then unitarily transformed to the rotating frame:

$$|\Psi_{cr}\rangle = e^{-i(\omega_{cr} L + E) t / \hbar} |\Phi_{cr}\rangle \qquad (3)$$

to obtain the stationary cranking model equation:

$$H_{cr} \cdot \bar{\Phi}_{cr} \equiv (H - \omega_{cr} \cdot L) \cdot \bar{\Phi}_{cr} = \bar{E}_{cr} \bar{\Phi}_{cr} \qquad (4)$$

where $L$ is the angular momentum operator. The angular frequency $\omega_{cr}$ is then determined by requiring $\Psi_{cr}$ or $\Phi_{cr}$ to have a given value $J$ of the angular momentum:

$$\hbar J \equiv \langle \Phi_{cr} | L | \Phi_{cr} \rangle \qquad (5)$$

The energy $E_{cr}$ in a space-fixed frame is then given by:

$$E_{cr} = \langle \Phi_{cr} | H | \Phi_{cr} \rangle = \langle \Phi_{cr} | H_{cr} + \omega_{cr} \cdot L | \Phi_{cr} \rangle = \bar{E}_{cr} + \omega_{cr} \cdot \langle \Phi_{cr} | L | \Phi_{cr} \rangle \qquad (6)$$

The physical or dynamical moment of inertia $\mathcal{I}_d$ is then given, at each value $J$, by the excitation energy $\Delta E_J$:

$$\frac{2\mathcal{I}_d}{\hbar^2} = \frac{4J-2}{\Delta E_J - \Delta E_{J-2}} \quad (MeV)^{-1} \qquad (7)$$

$$\Delta E_J \equiv E_J - E_0 \qquad (8)$$

Because in the cranking model the rotation is externally driven, the model is semi-classical and phenomenological in nature, and Eq. (4) is time-reversal non-invariant. It is desirable to have a model where the rotation is driven by the motions of the nucleons themselves instead of externally, i.e., it is desirable to derive the model microscopically, as suggested in [2,4,5,7, 12,18]. In several studies starting from first principles, Eq. (4) was derived using canonical transformation, projection and generator-coordinate methods using various approximations such as redundant coordinates, large deformations, expansion in power of the angular momentum, etc. [4,5,12,18, 27-30].

In this article, we derive the conventional cranking-model Schrodinger equation from first principles in a relatively simple manner using a canonical transformation to a rotating frame defined by the motions of the nucleons themselves without imposing any constraints on the wavefunction or the particle coordinates. The resulting microscopic cranking-model Schrodinger equation resembles closely Eq. (6), with a rotation angular velocity that depends on the nucleon coordinates, and with a total angular momentum that is the sum of those of the intrinsic system and rotating frame similar to that in the phenomenological nuclear rotor model and generalizing Eq. (5). In this article, the rotating-frame rotation is chosen to have a combined rigid and irrotational flow character. The functional form of the kinematic moment of inertia associated with the rotating frame depends on the nature of this combination.

In Section 2, we present the derivation of the microscopic cranking-model Schrodinger equation and compare it to the conventional cranking model. In Section 3, we solve the microscopic cranking-model Schrodinger equation using a self-consistent single-particle mean-field deformed harmonic oscillator potential, and compare the results with those of the conventional cranking model and experiments for the nucleus $^{20}_{10}Ne$. Section 4 presents concluding remarks.

## 2. Derivation of microscopic cranking model

The microscopic cranking model is derived by transforming the nuclear Schrodinger equation (instead of the Hamiltonian) to a reference frame rotating about the $x$ or 1 axis, defined by the collective Euler angle $\theta(x_{nj})$ and using the product wavefunction similar to that in [31]:

$$\Psi = G(\theta) \cdot \Phi(x_{nj}) \tag{9}$$

where $x_{nj}$ ($n = 1, ..., A;\ j = 1, 2, 3,$ where $A$ = nuclear mass number) are the space-fixed nucleon co-ordinates. Applying $\dfrac{\partial}{\partial x_{nj}}$ and $\dfrac{\partial^2}{\partial x_{nj}^2}$ to $\Psi$ in Eq. (9), we obtain:

$$\frac{\partial \Psi}{\partial x_{nj}} = \frac{\partial G}{\partial x_{nj}} \Phi + G \frac{\partial \Phi}{\partial x_{nj}} \tag{10}$$

$$\begin{aligned}
\frac{\partial^2 \Psi}{\partial x_{nj}^2} &= \frac{\partial^2 G}{\partial x_{nj}^2} \Phi + 2 \frac{\partial G}{\partial x_{nj}} \cdot \frac{\partial \Phi}{\partial x_{nj}} + G \frac{\partial^2 \Phi}{\partial x_{nj}^2} \\
&= \Phi \frac{\partial \theta}{\partial x_{nj}} \cdot \frac{\partial}{\partial \theta} \left( \frac{\partial \theta}{\partial x_{nj}} \cdot \frac{\partial G}{\partial \theta} \right) + 2 \frac{\partial \theta}{\partial x_{nj}} \cdot \frac{\partial G}{\partial \theta} \cdot \frac{\partial \Phi}{\partial x_{nj}} + G \frac{\partial^2 \Phi}{\partial x_{nj}^2}
\end{aligned} \tag{11}$$

Substituting Eq. (11) into the Schrodinger equation:

$$H \cdot \Psi \equiv \left( \frac{1}{2M} \sum_{n,j=1}^{N,3} p_{nj}^2 + V \right) \cdot \Psi = E \Psi \tag{12}$$

we obtain:

$$G \cdot H \cdot \Phi - \frac{\hbar^2}{M} \cdot \frac{\partial \theta}{\partial x_{nj}} \cdot \frac{\partial G}{\partial \theta} \cdot \frac{\partial \Phi}{\partial x_{nj}} - \frac{\hbar^2}{2M} \cdot \Phi \frac{\partial \theta}{\partial x_{nj}} \cdot \frac{\partial}{\partial \theta} \left( \frac{\partial \theta}{\partial x_{nj}} \cdot \frac{\partial G}{\partial \theta} \right) = E \Phi \quad (13)$$

Because the rotation of the frame through the angle $\theta$ is generated by the particle motions, we can consider the angular momentum operator component $L$ along $x$ or 1 axis and $\theta$ to be a canonically conjugate pair, satisfying the commutation relation:

$$[\theta, L] = i\hbar \quad \Rightarrow L \equiv \sum_n (y_n p_{nz} - z_n p_{ny}) = -i\hbar \frac{\partial}{\partial \theta} \quad (14)$$

Substituting Eq. (14) into Eq. (13), we obtain:

$$G \cdot H \cdot \Phi + \frac{1}{M} \cdot \sum_{nj} \frac{\partial \theta}{\partial x_{nj}} \cdot (L \cdot G) \cdot p_{nj} \Phi + \frac{1}{2M} \cdot \Phi \sum_{nj} \frac{\partial \theta}{\partial x_{nj}} \cdot L \cdot \left( \frac{\partial \theta}{\partial x_{nj}} \cdot L \cdot G \right) = E \Phi \quad (15)$$

Next we assume that $G$ is an eigenstate of $L$:

$$L e^{i\gamma\theta} = \hbar \gamma e^{i\gamma\theta} \quad (16)$$

where $\hbar\gamma$ is the angular momentum associated with the rotating frame. $\gamma$ is determined in Subsections 2.1 and 2.2. Substituting Eq. (16) into Eq. (15), we obtain:

$$\begin{aligned} H \cdot \Phi + \frac{\hbar\gamma}{M} \cdot \sum_{nj} \frac{\partial \theta}{\partial x_{nj}} \cdot p_{nj} \cdot \Phi + \frac{\hbar^2 \gamma^2}{2M} \cdot \sum_{nj} \frac{\partial \theta}{\partial x_{nj}} \cdot \frac{\partial \theta}{\partial x_{nj}} \cdot \Phi \\ + \frac{\hbar^2 \gamma^2}{4M} \cdot \Phi \cdot \sum_{nj} \cdot L \cdot \left( \frac{\partial \theta}{\partial x_{nj}} \cdot \frac{\partial \theta}{\partial x_{nj}} \right) = E \Phi \end{aligned} \quad (17)$$

We now choose $\theta$ to satisfy the relation (for rotation about $x$ or 1 axis only):

$$\frac{\partial \theta}{\partial x_{nj}} = \sum_{k=1}^{2} \chi_{jk} x_{nk}, \quad \chi_{jk} = 0 \text{ for } j, k \neq 2, 3 \quad (18)$$

The 3x3 matrix $\chi$ can be chosen to be a sum of different types of matrices, each describing different types of physical motions such as rigid, irrotational, and non-quadrupole type flow regimes described in [32-34]. These options ise explored in more detail in part II of this article. In this article, we choose $\chi$ to be the sum of a symmetric and an antisymmetric matrices so that the non-zero elements of $\chi$ are $\chi_{23} \equiv \chi_2 + \chi_3$, and $\chi_{32} = -\chi_2 + \chi_3$. We choose $\chi_3 = \lambda \cdot \chi_2$ and hence:

$$\chi_{23} \equiv (1+\lambda) \cdot \chi_2, \text{ and } \chi_{32} = -(1-\lambda) \cdot \chi_2 \tag{19}$$

Substituting Eq. (19) into $[\theta, L] = i\hbar$ in Eq. (14), we obtain:

$$\chi_2 = -(\mathcal{J}_+ - \lambda \cdot \mathcal{J}_-)^{-1} \equiv -\mathcal{J}^{-1} \tag{20}$$

where:

$$\mathcal{J}_+ \equiv \sum_n (y_n^2 + z_n^2), \quad \mathcal{J}_- \equiv \sum_n (y_n^2 - z_n^2) \tag{21}$$

Substituting Eqs. (18) and (19) into Eq. (17), we obtain:

$$\left[ H + \frac{\hbar\gamma}{M\mathcal{J}} \cdot (L - \lambda \cdot T) + \frac{\hbar^2 \gamma^2}{2M\mathcal{J}} \cdot \left( 1 + \lambda^2 - \lambda \frac{1-\lambda^2}{\mathcal{J}} \cdot \mathcal{J}_- \right) \right. $$
$$\left. + \frac{2i\hbar^2 \gamma \lambda^2}{M\mathcal{J}^2} \cdot \left( \lambda - \frac{1-\lambda^2}{\mathcal{J}} \cdot \mathcal{J}_- \right) \cdot \sum_n y_n z_n \right] \cdot \Phi = E\Phi \tag{22}$$

where $T$ is a linear shear operator, generating a linear irrotational flow, and defined by:

$$T \equiv \sum_n (y_n p_{nz} + z_n p_{ny}) \tag{23}$$

We note that the second term on the left-hand-side of Eq. (22) is the cranking or Coriolis energy term. This term comes from the second term on the left-hand-side of Eq. (17), which represents the interaction between the rotating frame and intrinsic motion. The remaining terms on the left-hand-side of Eq. (22) arise from centrifugal stretching effect and fluctuations in the rotating-frame angular velocity. These remarks and those in [35] hopefully provide better understanding of the forces involved in the rotational motion.

## 2.1 Rigid-flow rotating frame

For $\lambda = 0$, Eq. (22) reduces to:

$$\left( H + \frac{\hbar\gamma}{M\mathcal{J}_+} \cdot L + \frac{\hbar^2 \gamma^2}{2M\mathcal{J}_+} \right) \cdot \Phi = E\Phi \tag{24}$$

where $M\mathcal{J}_+$ is the rigid-flow moment of inertia and commutes with $L$. Expressed in terms of the rigid-flow angular frequency $\omega_{rig}$ of the rotating frame:

$$\omega_{rig} \equiv \frac{\hbar\gamma}{M\mathcal{J}_+}, \tag{25}$$

Eq. (24) becomes:

$$\left(H + \omega_{rig} \cdot L + \frac{1}{2}\omega_{rig}^2 M \mathcal{I}_+\right) \cdot \Phi = E\Phi \tag{26}$$

Eq. (26) resembles the conventional cranking model equation in a space-fixed frame given in Eq. (6) but differs from Eq. (6) by the rigid-flow kinetic energy term (third term on the left-hand-side of Eq. (26)) and by the microscopically defined $\omega_{rig}$ instead of the constant parameter $\omega_{cr}$ in Eq. (6). Eq. (26) also differs from the cranking model in the rotating frame in Eq. (4) by the aforementioned terms and by the sign of $\omega_{rig}$ term (note that the microscopic cranking model solves Eq. (24) whereas the conventional cranking model solves Eq. (4) and not Eq. (6)). Eq. (26) is time-reversal invariant whereas Eq. (6) is not.

Another difference is that, in the microscopic cranking model, the rotating frame angular momentum $\hbar\gamma$ in Eq. (25) is determined by requiring the expectation of the total angular momentum operator $L$ with respect to the wavefunction $\Psi$ in Eq. (9) to have the experimentally observed rotational-band excited-state angular momentum value $\hbar J$ [1]:

$$\hbar J = \langle \Psi | L | \Psi \rangle = \langle G | L | G \rangle + \langle \Phi | L | \Phi \rangle = \hbar\gamma + \hbar l = \omega_{rig} \cdot M \mathcal{I}_+ + \hbar l \tag{27}$$

where:

$$\hbar l \equiv \langle \Phi | L | \Phi \rangle \tag{28}$$

The prescription in Eq. (27) differs from that in Eq. (5) for the conventional cranking model by the angular momentum $\hbar\gamma$ of the rotating frame.

It follows from Eq. (27) that, in the microscopic cranking model, $\gamma$ (and hence $\omega_{rig}$) and $l$ can have different signs, and hence the rotating frame and the intrinsic system may rotate in opposite directions, unlike that in the conventional cranking model but similarly to that in the phenomenological nuclear rotor model [4,6,12,35].

The excitation energy and physical or dynamic moment of inertia are defined in Eqs. (7) and (8). The rigid-flow moment of inertia $\mathcal{I}_+$ in Eqs. (24)-(26) may be called the kinematic moment of inertia.

The above-stated differences between the microscopic and conventional cranking models generate some significant differences in the predictions of the two models as is demonstrated in Section 4.

---

[1] The value of $\gamma$ determined by Eq. (27) is an approximation to the integer value of $\gamma$ needed to ensure that $\Psi$ is single-valued.

## 2.2 Rigid-plus-irrotational flow rotating frame

For $\lambda \neq 0$, Eq. (22) generalizes the microscopic cranking model Eq. (24) or (26) to include linear shear flow kinematics. Because $\lambda$ is expected to be less than unity, the kinematic moment of inertia $M\mathcal{I}$ is expected to be relatively large, and the off-diagonal element of the quadrupole tensor $\sum_n y_n z_n$ is expected to be relatively small, we may ignore most of the terms in the third and fourth terms on the left-hand-side in Eq. (22). In fact, if we choose $\lambda$ such that:

$$\lambda \cdot \mathcal{I} - (1 - \lambda^2) \cdot \mathcal{I}_- = 0 \tag{29}$$

all the $\lambda$-dependent quantities in the brackets in the third and fourth terms on the left-hand-side of Eq. (22) would vanish identically[2]. Therefore, in this article, we ignore these terms, and Eq. (22) reduces to:

$$\left[ H + \omega_{rs} \cdot (L - \lambda \cdot T) + \frac{1}{2} \omega_{rs}^2 M \mathcal{I} \right] \cdot \Phi = E \Phi \tag{30}$$

where the rigid-plus-shear angular frequency $\omega_{rs}$ of the rotating frame is defined by:

$$\omega_{rs} \equiv \frac{\hbar \gamma}{M \mathcal{I}}, \tag{31}$$

The order of the appearance of the operators in the second term on the left-hand-side in Eq. (30) is immaterial because we can readily show that:

$$[\mathcal{I}, L - \lambda \cdot T] = 0 \tag{32}$$

for any c-number $\lambda$.

The excitation energy and physical or dynamic moment of inertia are defined in Eqs. (7) and (8). The variable $\mathcal{I}$ in Eqs. (30) and (31) may be called the kinematic moment of inertia.

As in the rigid-flow rotating-frame case in Eq. (27), the rotating frame angular momentum $\hbar \gamma$ in Eqs. (30) and (31) is determined by requiring the expectation of the total angular momentum operator $L$ with respect to the wavefunction $\Psi$ in Eq. (9) to have the experimentally observed rotational-band excited-state angular momentum value $\hbar J$ [3]:

---

[2] Substituting the expression in Eq. (20) for $\mathcal{I}$ into Eq. (29) and solving the resulting equation for $\lambda$, we obtain $\lambda = \mathcal{I}_- / \mathcal{I}_+$ yielding $\mathcal{I} = \mathcal{I}_+ - \mathcal{I}_-^2 / \mathcal{I}_+$, which is the rigid-flow moment of inertia reduced by the irrotational-flow moment of inertia.

[3] The value of $\gamma$ determined by Eq. (25) is an approximation to the integer value of $\gamma$ needed to ensure that $\Psi$ is single-valued.

$$\hbar J = \langle \Psi | L | \Psi \rangle = \langle G | L | G \rangle + \langle \Phi | L | \Phi \rangle = \hbar \gamma + \hbar l = \omega_{rs} \cdot M \mathcal{J} + \hbar l \tag{33}$$

where:

$$\hbar l \equiv \langle \Phi | L | \Phi \rangle \tag{34}$$

The prescription in Eq. (33) differs from that in Eq. (5) for the conventional cranking model by the angular momentum $\hbar \gamma$ of the rotating frame.

It follows from Eq. (33) that, in the microscopic cranking model, $\gamma$ (and hence $\omega_{rs}$) and $l$ can have different signs, and hence the rotating frame and the intrinsic system may rotate in opposite directions, unlike that in the conventional cranking model.

The above-stated differences between the microscopic and conventional models generate some significant differences in the predictions of the two models as is demonstrated in Section 4.

### 3. Solutions of Eqs. (4), (26), and (30)

In this section, we determine the solutions of Eqs. (4), (26), and (30) for a mean-field deformed harmonic oscillator Hamiltonian:

$$H = \frac{1}{2M} \sum_{n,j=1}^{N,3} p_{nj}^2 + \frac{M\omega_1^2}{2} \sum_n x_n^2 + \frac{M\omega_2^2}{2} \sum_n y_n^2 + \frac{M\omega_3^2}{2} \sum_n z_n^2 \tag{35}$$

### 3.1 Solution of conventional cranking model Eq. (4)

The solution to Eq. (4) has been obtained by a number of authors [36-40] using a canonical or unitary transformation to eliminate the cross terms $y_n p_{nz}$ and $z_n p_{ny}$, and obtain the transformed harmonic oscillator Hamiltonian:

$$\bar{H}_{cr} = \frac{1}{2M} \sum_{n,j=1}^{N,3} p_{nj}^2 + \frac{M\omega_1^2}{2} \sum_n x_n^2 + \frac{M\alpha_2^2}{2} \sum_n y_n^2 + \frac{M\alpha_3^2}{2} \sum_n z_n^2 \tag{36}$$

and the energy eigenvalue in the rotating frame:

$$\bar{E}_{cr} = \hbar \omega_1 \Sigma_1 + \hbar \alpha_2 \Sigma_2 + \hbar \alpha_3 \Sigma_3 \tag{37}$$

where:

$$\alpha_2^2 \equiv \omega_+^2 + \omega_{cr}^2 + \sqrt{\omega_-^4 + 8\omega_{cr}^2 \omega_+^2}, \qquad \alpha_3^2 \equiv \omega_+^2 + \omega_{cr}^2 - \sqrt{\omega_-^4 + 8\omega_{cr}^2 \omega_+^2} \tag{38}$$

$$\omega_+^2 \equiv \frac{\omega_2^2 + \omega_3^2}{2}, \qquad \omega_-^2 \equiv \frac{\omega_2^2 - \omega_3^2}{2}, \qquad \Sigma_k \equiv \sum_{n_k=0}^{n_{kf}} (n_k + 1/2) \tag{39}$$

The energy in the space-fixed frame is then given in Eq. (6).

For $\bar{H}_{cr}$ in Eq. (35) to approximate a Hartree-Fock mean-field Hamiltonian, the frequencies $\omega_k$ ($k = 1,2,3$) are chosen to minimize the energy $\bar{E}_{cr}$ in Eq. (37) with respect to $\omega_k$ at each fixed value of $J$ and hence of $\omega_{cr}$ given by the constraint in Eq. (5), and subject to a constant nuclear-quadrupole-volume condition:

$$\langle x^2 \rangle \cdot \langle y^2 \rangle \cdot \langle z^2 \rangle = c_o \qquad (40)$$

where $\langle x_k^2 \rangle \equiv \langle \bar{\Phi}_{cr} | \sum_n x_{nk}^2 | \bar{\Phi}_{cr} \rangle$. This minimization yields a self-consistency between the shapes of the nuclear equi-potential and equi-density surfaces [7,36-39]. The minimization stated above is performed numerically [38,39].

### 3.2 Solution of microscopic cranking Eqs. (26) and (30)

The variable $\mathcal{J}^{-1}$ in $\omega_{rs}$ in Eqs. (30) and (31) is a many body operator. Therefore, to find a solution of Eq. (30), we may regard $\mathcal{J}$ as an independent dynamical collective variable and transform Eq. (30) to $\mathcal{J}$ using the method suggested in [41-43]. This approach would account for the interaction of fluctuations in $\mathcal{J}$ with the rotation. In this article, however, we solve Eq. (30) by replacing $\mathcal{J}$ by its expectation value in the state $\Phi$:

$$\mathcal{J}^o \equiv \langle \Phi | \mathcal{J} | \Phi \rangle = \langle \Phi | \mathcal{J}_+ | \Phi \rangle - \lambda \langle \Phi | \mathcal{J}_- | \Phi \rangle \equiv \mathcal{J}_+^o - \lambda \mathcal{J}_-^o \qquad (41)$$

Thereby suppressing fluctuations in $\mathcal{J}$ and $\omega_{rs}$ in Eq. (31) and their interaction with the collective rotation. This approximation is justified because $\mathcal{J}^o$ is relatively large ($\sim 35\hbar$) and varies gradually from state to state. Eqs. (30) and (31) then become:

$$\left[ H + \bar{\omega}_{rs} \cdot (L - \lambda \cdot T) + \frac{1}{2} \bar{\omega}_{rs}^2 M \mathcal{J}^o \right] \cdot \Phi = E \Phi \qquad (42)$$

$$\bar{\omega}_{rs} \equiv \frac{\hbar \gamma}{M \mathcal{J}^o}, \qquad (43)$$

From the statement in footnote 1, $\lambda$ may then be considered to be a function of $\mathcal{J}_-^o / \mathcal{J}_+^o$. We may also determine $\lambda$ from a minimization of the energy $E$ in Eq. (42) with respect to $\lambda$.

The solution of Eq. (42) is obtained similarly to that of Eq. (4) in Subsection 3.1 (and specialized to the solution of Eq. (26) by setting $\lambda$ to zero). We then obtain the transformed harmonic oscillator Hamiltonian:

$$\bar{H} = \frac{1}{2M} \sum_{n,j=1}^{N,3} p_{nj}^2 + \frac{M\omega_1^2}{2} \sum_n x_n^2 + \frac{M\alpha_2^2}{2} \sum_n y_n^2 + \frac{M\alpha_3^2}{2} \sum_n z_n^2 \qquad (44)$$

and the energy eigenvalue in the rotating frame:

$$\bar{E} \equiv E - \frac{1}{2}\bar{\omega}_{rs}^2 M \mathcal{J}^o = \hbar \omega_1 \Sigma_1 + \hbar \alpha_2 \Sigma_2 + \hbar \alpha_3 \Sigma_3 \qquad (45)$$

where:

$$\alpha_2^2 \equiv \omega_+^2 + (1-\lambda^2)\cdot \bar{\omega}_{cr}^2 + \sqrt{\omega_-^4 + 8\omega^2 \cdot \omega_\lambda^2}, \quad \alpha_3^2 \equiv \omega_+^2 + (1-\lambda^2)\cdot \bar{\omega}_{cr}^2 + \sqrt{\omega_-^4 + 8\omega^2 \cdot \omega_\lambda^2} \qquad (46)$$

where $\Sigma_k$ are defined in Eq. (39), and:

$$\omega_\lambda^2 \equiv \omega_+^2 + \lambda \cdot \omega_-^2 \qquad (47)$$

The energy in the space-fixed frame is then given by:

$$E = \frac{1}{2}\bar{\omega}_{rs}^2 M \mathcal{J}^o + \hbar \omega_1 \Sigma_1 + \hbar \alpha_2 \Sigma_2 + \hbar \alpha_3 \Sigma_3 \qquad (48)$$

For $\bar{H}$ in Eq. (44) to approximate a Hartree-Fock mean-field Hamiltonian, the frequencies $\omega_k$ ($k=1,2,3$) are chosen to minimize the energy $E$ in Eq. (48) with respect to $\omega_k$ at each fixed value of $J$ and hence of $\bar{\omega}_{rs}$ and $\mathcal{J}^o$ given by the constraint in Eq. (33), and subject to a constant nuclear-quadrupole-volume condition:

$$\langle x^2 \rangle \cdot \langle y^2 \rangle \cdot \langle z^2 \rangle = c_o \qquad (49)$$

where $\langle x_k^2 \rangle \equiv \langle \Phi | \sum_n x_{nk}^2 | \Phi \rangle$. This minimization yields a self-consistency between the shapes of the nuclear equi-potential and equi-density surfaces [7,36-39]. The minimization stated above is performed numerically as in [38,39].

## 4. Model predictions for $^{20}_{10}Ne$

In this section we present the results of the application of the microscopic cranking model as described in Sections 2.1, 2.2 and 3.2 to the nucleus $^{20}_{10}Ne$ and compare these results with those of the conventional cranking model in Sections 1 and 3.1.

For $^{20}_{10}Ne$, we use the deformed harmonic oscillator nucleon-occupation configuration $(\Sigma_1, \Sigma_2, \Sigma_3) = (14,14,22)$, with the spherical harmonic oscillator frequency $\hbar \omega_o = 35.4 \cdot A^{-1/3}$ $MeV$ as in [37,38].

## 4.1 Rigid-flow rotating frame

For the rigid-flow rotating-frame case described in Section 2.1 (i.e. for $\lambda = 0$), we have performed model calculations using various configurations for $^{20}_{10}Ne$ to learn how the model behaves. We present the results of this survey and possible physical explanations for the results.

Fig 1 compares the excitation energy $\Delta E_J$ observed empirically and predicted by the microscopic and conventional cranking models when the $\omega_k$'s are kept constant at their ground-state (i.e., $J = 0$) self-consistent values (i.e., the nucleus remains prolate at all $J$ values). Fig 1 shows that the microscopic cranking model predicts well the excitation energy whereas the conventional cranking model predicts a lower $\Delta E_J$. The excitation energy in all cases increases with $J$ except at $J = 8$, the empirically observed $\Delta E_J$ is significantly lower than that predicted by either of the models. This difference may be attributed to a Coriolis-force induced quasi-particle re-alignment not included in the models. We note that in this case the rotational band does not terminate at $J = 8$ in either the microscopic or conventional cranking model.

Fig 2 shows that the measured quadrupole moment $Q_o$ decreases monotonically with $J$ and there is a sharp drop in $Q_o$ at $J = 8$ when the nucleus presumably becomes symmetric about the rotation axis (i.e., when $\omega_2$ and $\omega_3$ become equal). The microscopic and conventional cranking models predict lower $Q_o$, which decreases up to $J = 5$ in the microscopic and up to $J = 8$ in the conventional cranking models, and increases above these $J$ values (the increase in $Q_o$ for the cranking model is not included in Fig 2).

The intrinsic angular momentum predicted by the microscopic model decreases from about $-5\hbar$ at $J = 2$ to $-9\hbar$ at $J = 8$, and in this range of $J$, the angular momentum of the rotating frame increases from $8\hbar$ to $16.5\hbar$. The predicted rotational band terminates above $J = 23$ when the intrinsic oscillator frequency $\alpha_3$ vanishes and hence the model governing equations have no solutions. These results indicate that, to predict the observed change in the nucleus shape (i.e., $Q_o$) the microscopic cranking model must account for nuclear volume conservation at all $J$ values.

Figs 3 and 4 show respectively the excitation energies and quadrupole moment when we use $\omega_k$'s determined self-consistently at all $J$ values. The conventional cranking model predicts higher $\Delta E_J$ above $J = 4$, and a $Q_o$ that decreases similarly to the observed $Q_o$ up to $J = 8$, at which point the nucleus becomes axially symmetric about the rotation axis (i.e., $\omega_2$ and $\omega_3$ become equal) and the rotational band terminates. The microscopic model predicts significantly higher $\Delta E_J$, and a $Q_o$ that decreases rapidly up to $J = 5$ and increases slightly thereafter.

Fig 5 and 6 show respectively the excitation energies and quadrupole moment for the nucleon-occupation configuration changed to $(\Sigma_1, \Sigma_2, \Sigma_3) = (14,14,16)$. The microscopic model predicts quite well $\Delta E_J$. The model predicted much lower $Q_o$, as expected, but it decreases gradually with $J$. The intrinsic angular momentum is nearly constant at $-2\hbar$ and the rotating-frame angular momentum increases from $3\hbar$ to $10\hbar$.

The differences between the results presented in Figs 1 and 2 and those in Figs 3 and 4 (and also those in Figs 5 and 6) seem to indicate that in the cranking model the values of $\Delta E_J$ and $Q_o$ are determined significantly more by changes in the nuclear shape (i.e., by changes in $\omega_k$'s of the potential energy) than by changes in the kinetic energy.

The results presented above show that the microscopic cranking model predicts somewhat similar results to those of the conventional cranking model. However, in the microscopic cranking model, the rotations of the intrinsic system and rotating frame are in opposite directions, and the intrinsic system angular momentum reaches its limiting value at lower $J$ value than in the conventional cranking model.

## 4.2 Rigid-plus-irrotational flow rotating frame

For the rigid-plus-linear irrotational flow case described in Section 2.2 (i.e. for $\lambda \neq 0$), the model predicts results very close to those of the rigid-flow case in Subsection 4.1. Therefore, adding linear irrotational flow rotation component to the rigid-flow rotation of the rotating frame does not have a significant effect on the results of the rigid-flow case. This weak impact on the model results arises from near cancellation of the $\lambda$-dependent terms in the model governing equations.

Therefore, in part II of this article, we study the impact of other flow regimes, such as the non-quadrupole rotation, to obtain better agreement with experiment.

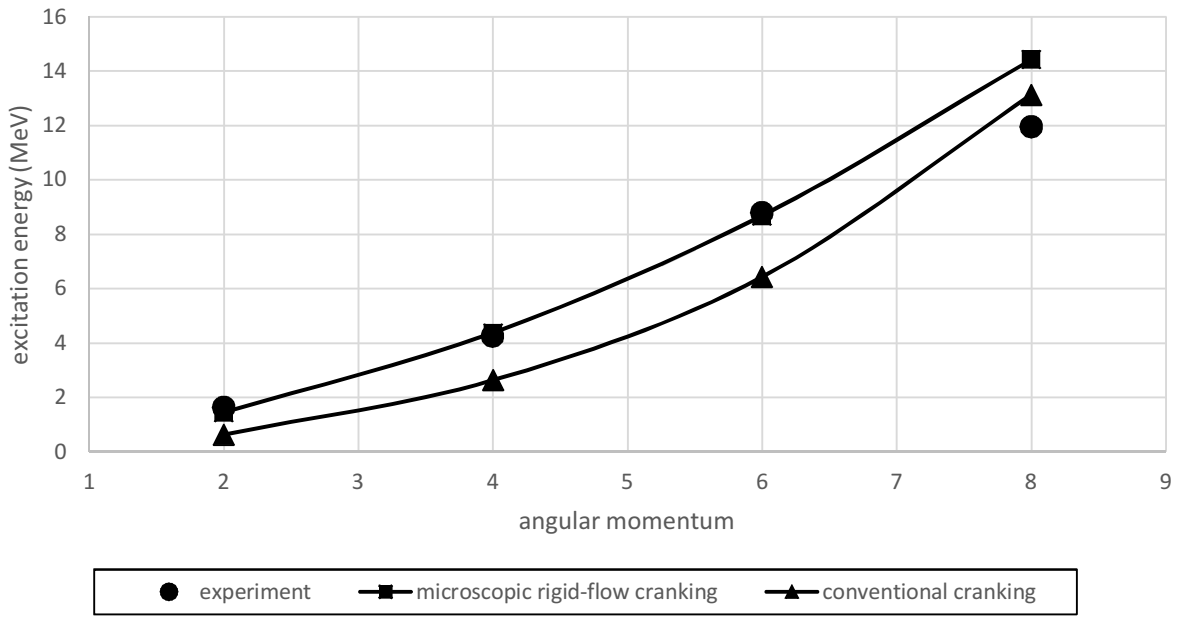

Fig 1: excitation energy versus J for self-consistency at J = 0 only

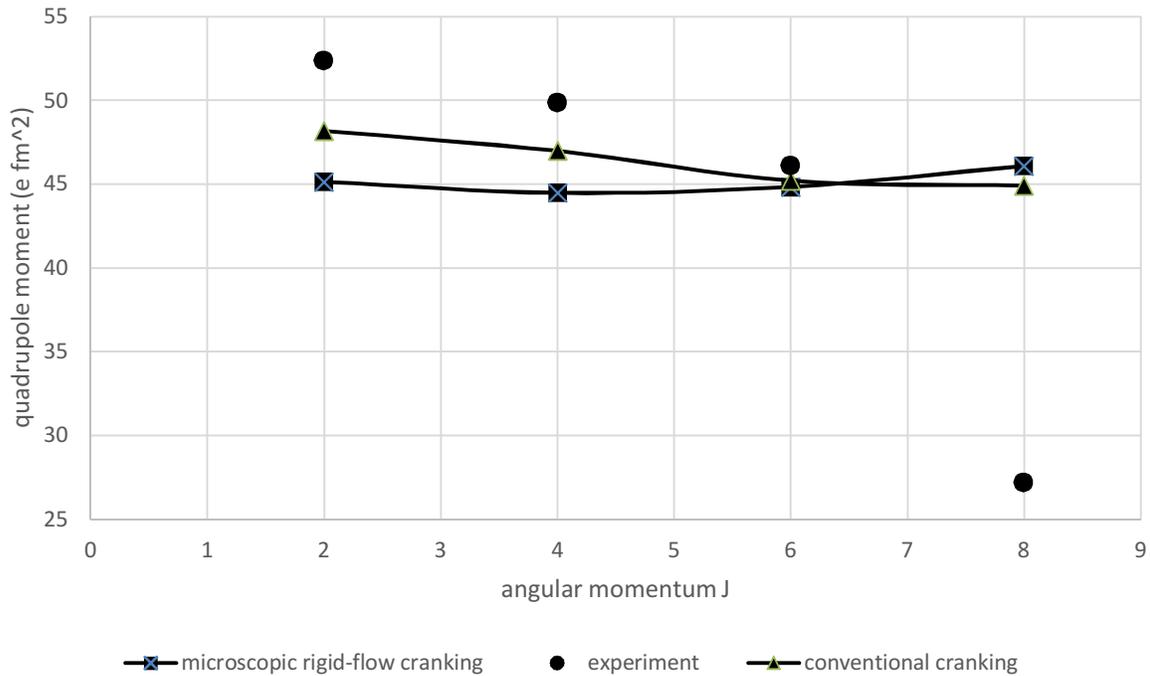

Fig 2: quadrupole moment versus $J$ for self-consistency at $J = 0$ only

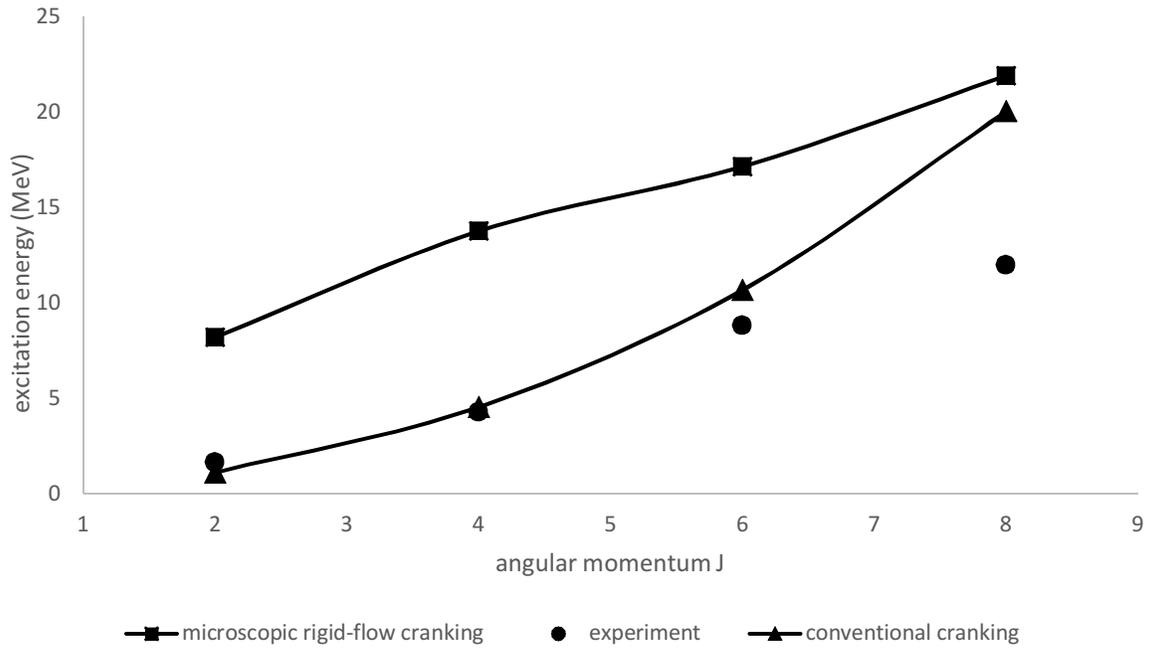

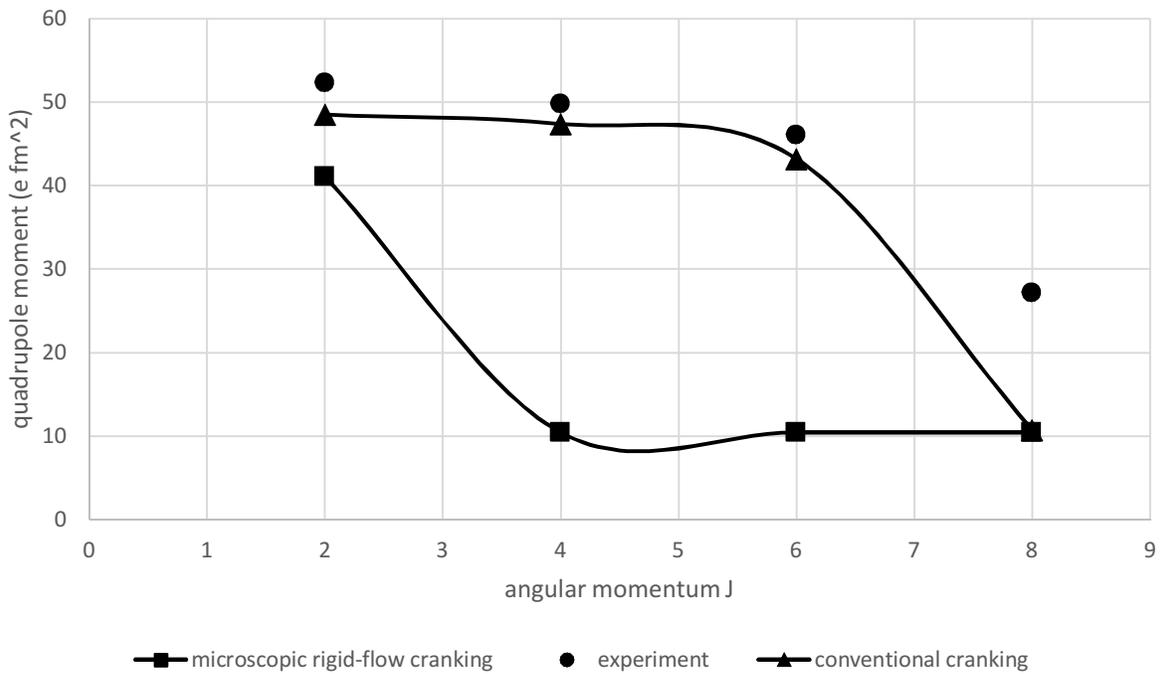

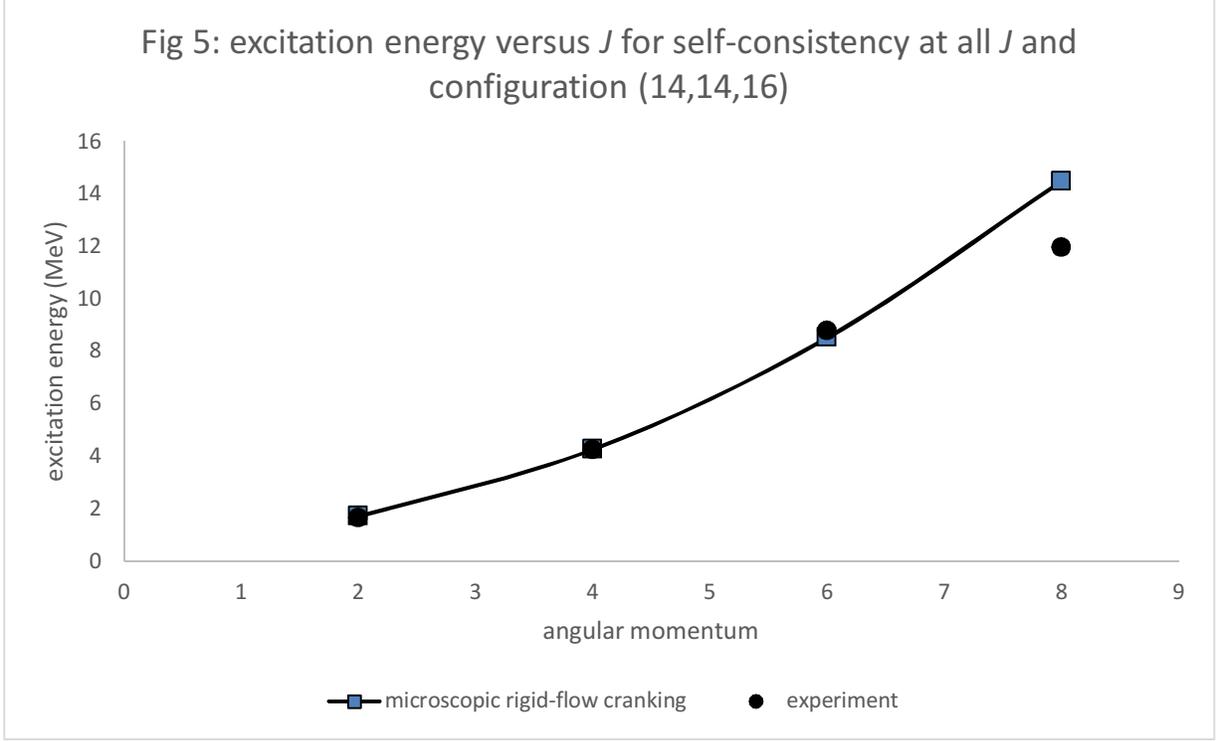

Fig 5: excitation energy versus *J* for self-consistency at all *J* and configuration (14,14,16)

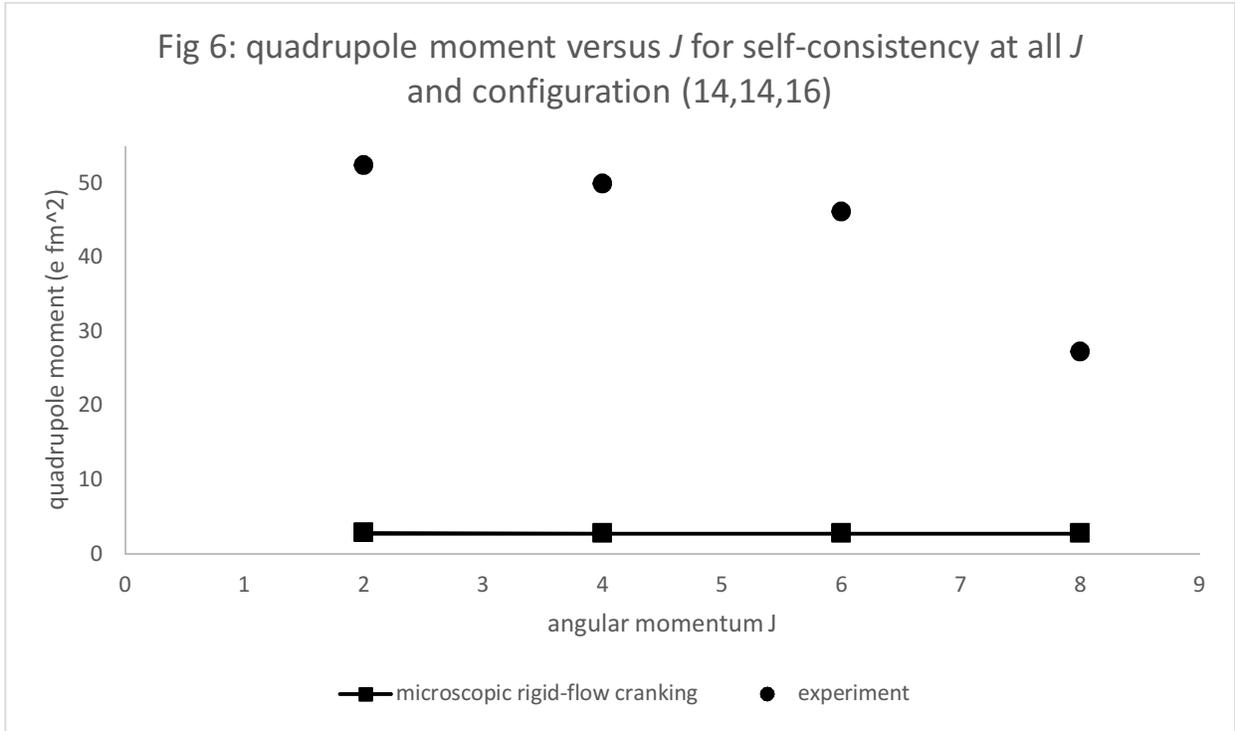

Fig 6: quadrupole moment versus *J* for self-consistency at all *J* and configuration (14,14,16)

## 5. Concluding remarks

The phenomenological semi-classical conventional cranking model is often used to study rotational properties of nuclei. In view of its importance in nuclear structure studies it is desirable to have a better understanding of the assumptions and approximations that underlie the model derivation. In the hope of achieving this objective, we attempt, in this article, to derive the model in a relatively simple manner from first principles. We simply transform the nuclear Schrodinger equation (instead of the Hamiltonian) to a rotating reference frame using a product wavefunction and imposing no constraints on either the wavefunction or the particle coordinates. The rotation of the reference frame is generated by the motions of the nucleons, instead of being imposed externally as in the conventional cranking model. As a consequence, the microscopic model Schrodinger equation is time reversal invariant unlike the conventional model. In this article, the rotation of the reference frame is chosen to be determined by a combination of rigid and irrotational flows.

The resulting transformed Schrodinger equation resembles that of the conventional cranking model in a space-fixed frame. However, the angular velocity in the microscopic model is not an constant parameter but is a dynamical variable determined by angular momentum of the rotating frame and a kinematic moment of inertia the nature of which is determined by the aforementioned flow combination. The angular momentum of the rotating frame is determined by equating the angular momentum of a member state of a rotational-band to the predicted expectation value of the total angular momentum. This expectation is a sum of the angular momenta of the rotating frame and intrinsic system. It turns out that the rotation of the intrinsic system and rotating frame are in the opposite directions. Furthermore, the transformed Schrodinger equation has, in addition to the Coriolis energy term, a rigid-flow type kinetic energy term that is absent from the conventional cranking model Schrodinger equation.

In this article, we suppress the fluctuations in the angular velocity and their coupling to the intrinsic motion by replacing the kinematic moment of inertia by its expectation value in each angular momentum state. This approximation is justified because the impact of the fluctuations can be shown to be small. The resulting Schrodinger equation is then solved for a deformed harmonic oscillator mean-field potential. The oscillator frequencies are determined self-consistently from numerical energy minimization subject to constant nuclear volume condition.

For the nucleus $^{20}_{10}Ne$ the excitation energy and quadrupole moment are calculated for the ground-state rotational band with various nucleon configurations with and without self-consistency. The results show that the model predictions have trends similar to those of the conventional cranking model except for the intrinsic angular momentum, which has a negative value and reaches its limiting value at a lower angular momentum. The predicted excitation energy is higher than that observed in the experiment except in the case of a configuration with reduced deformation and in the case where the self-consistency is restricted to $J = 0$ state only. The predicted quadrupole moment is lower than that observed in the experiment. The irrational flow component in the rotation of the rotating frame is found to have small effect on the results.

The above results seem to indicate that rigid-flow prescription for the rotating frame may not be adequate and we need to include also other types of flows such as non-quadrupole type. These flows will be considered in part II of this article.

**References**

[1] D. R. Inglis, Phys. Rev. 103 (1956) 1786.

[2] D.J. Thouless and J.G. Valatin, Nucl. Phys. 31 (1962) 211.

[3] J.M. Eisenberg and W. Greiner, Nuclear Theory (North-Holland, Amsterdam, 1970).

[4] D.J. Rowe, Nuclear Collective Motion (Mathuen and Co. Ltd., London, 1970).

[5] R.A. Sorensen, Rev. Mod. Phys. 45 (1973) 353.

[6] A. deShalit and H. Feshbach, Theorectical Nuclear Physics, Vol. 1 (John Wiley & Sons, Inc., N.Y., 1974).

[7] A. Bohr and B.R. Mottelson, Nuclear Structure, Vol. II (Benjamin, N.Y., 1975).

[8] G, Ripka and J.P. Blaizot, in Heavy-Ion, High-Spin States and Nuclear Structure, Vol. 1, Trieste Int. seminar on nuclear physics, (IAEA, Vienna, 1975).

[9] R.M. Lieder and H. Ryde. in Advances in nuclear physics. Vol. 10, edited *by* M. Baranger and E. Vogt. Plenum Press, New York. 1978.

[10] R. Bengtsson and S. Frauendorf, Nucl. Phys. A **314**, (1979) 27.

[11] W.D. Heiss and R.G. Nazmitdinov, Phys. Letts. B 397 (1997) 1.

[12] P. Ring and P. Schuck, The Nuclear Many-Body Problem (Springer-Verlag, N.Y., 1980).

[13] A.L. Goodman, G.S. Goldhaber, A. Klein, and R.A. Sorensen, Nucl. Phys. A347 (1980) 1-428 (Proc. Int. Conf. on Bands Structure and Nuclear Dynamics, edited by, Tulane University, New Orleans, February 28-March 1980).

[14] A. Faessler, in Proc of Conf. on High Angular Momentum Properties of Nuclei, Proc. Oak Ridge, Tennessee, November 2-4, 1982, Vol. 4, edited by N.R. Johnson, Hardwood Academic Publishers N.Y. 1982.

[15] F.S. Stephens. in Frontiers in nuclear dynamics. edited by R.A. Broglia and C.H. Daso. Plenum Press, New York. 1985.

[16] S.G. Nilsson and I. Ragnarsson. Shapes and shells in nuclear structure (Cambridge University Press, Cambridge, UK. 1995).

[17] T. Tanaka, F. Sakata, T. Marumori, and K. Iwasawa. Phys. Rev. C, 56 (1997) 180.

[18]  A. Klein, Phys. Rev. C 63 (2000) 014316.

[19]  S. Frauendorf, Rev. Mod. Phys. 73 (2001) 463.

[20]  W.D. Heiss and R.G. Nazmitdinov, Phys. Rev. C 65 (2002) 054304.

[21]  R.G. Nazmitdinov, D. Almehed, F. Donau, Phys. Rev. C 65 (2002) 041307.

[22]  M. Matsuzaki, Y.R. Shimizu, and K. Matsuyanagi, Phys. Rev. C 65 (2004) 034325.

[23]  M.A. Deleplanque, S. Frauendorf, V.V. Pashkevich, S.Y. Chu, and A. Unzhakova. Phys. Rev. C, 69 (2004) 044309.

[24]  A.V. Afanasjev, arXiv [nucl-th] 1510.08400, 2015 Oct. 28.

[25]  A.G. Magner, D.V. Gorpinchenko, and J. Bartel, arXiv [nucl-th] 1604.06866, 2016 April 23.

[26]  T. Nakatsukasa, K. Matsuyanagi, M. Matsuzaki, Y.R. Shimizu, arXiv [nucl-th] 1605.01876, 2016 May 06.

[27]  R.E. Peierls and J. Yaccoz, Proc. Phys. Soc. 70 (1957) 381.

[28]  A. Bohr and B. Mottelson, Forh. Norske Vidensk. Selsk. 31 (1958) No. 12.

[29]  F. Villars and N. Schmeing-Rogerson, Ann. Phys. 63 (1971) 443.

[30]  F. Villars and G. Cooper, Ann. Phys. (N.Y.) 56 (1970) 224.

[31]  P. Gulshani, Nuclear Physics A 832, 18 (2010).

[32]  R.Y. Cusson, Nucl. Phys. A 114 (1968) 289.

[33]  P. Gulshani and D.J|. Rowe, Can. J. Phys. 54 (1976) 970.

[34]  B. Buck, L.C. Biedenharn, and R.Y. Cusson, Nucl. Phys. A 317 (1979) 205.

[35]  B.J. Verhaar, A.M. Schulte, and J. de Kam, Z. Physik A 277 (1976) 261.

[36]  J.G. Valatin, Proc. Roy. Soc. (London) 238 (1956) 132.

[37]  V.G. Zelevinskii, Sov. J. Nucl. Phys. 22 (1976) 565.

[38]  A.P. Stamp, Z. Physik A 284 (1878) 312.

[39]  T. Troudet and R. Arvieu, Ann. Phys. 134 (1981) 1.

[40]  P. Gulshani and A. B. Volkov, J. Phys. G: Nucl. Phys. 6 (1980) 1335.

[41]  P. Gulshani, arXiv [nucl-th] 1602.01324 , July 15, 2016.

[42]  P. Gulshani, Can. J. Phys. 93 (2015) 1.

[43]  P. Gulshani, arXiv [nucl-th] 1502.07590 , May 12, 2015.